# Shared Autonomous Vehicle Simulation and Service Design

Reza Vosooghi[a,b,*], Jakob Puchinger[a,b], Marija Jankovic[b], Anthony Vouillon[c]

[a]*Institut de Recherche Technologique SystemX, Palaiseau 91120, France*
[b]*Laboratoire Génie Industriel, CentraleSupeléc, Université Paris-Saclay, Gif-sur-Yvette 91190, France*
[c]*Direction de la Recherche / Nouvelles Mobilité (DEA-IRM), Technocentre Renault, Guyancourt 78280, France*

**Abstract**

Today, driverless cars, as a new technology that allows a more accessible, dynamic and intelligent form of Shared Mobility, are expected to revolutionize urban transportation. One of the conceivable mobility services based on driverless cars is shared autonomous vehicles (SAVs). This service could merge cabs, carsharing, and ridesharing systems into a singular transportation mode. However, the success and competitiveness of future SAV services depend on their operational models, which are linked intrinsically to the service configuration and fleet specification. In addition, any change in operational models will result in a different demand. Using a comprehensive framework of SAV simulation in a multi-modal dynamic demand system with integrated SAV user taste variation, this study evaluates the performance of various SAV fleets and vehicle capacities serving travelers across the Rouen Normandie metropolitan area in France. Also, the impact of ridesharing and rebalancing strategies on service performance is investigated.

Research results suggest that the performance of SAV is strongly correlated with the fleet size and the strategy of individual or shared rides. Further analysis indicates that for the pricing scheme proposed in this study (i.e., 20% lower for ridesharing scenario), the standard 4-seats car with shared ride remains the best option among all scenarios. The results also underline that enabling vehicle-rebalancing strategies may have an important effect on both user and service-related metrics. The estimated SAV average and maximum driven distance prove the importance of vehicle range and charging station deployment.

## 1. Introduction

Driverless cars or autonomous vehicles (AVs) represent a transformative technology that may have important implications for society, urban mobility, economy, and environment (Fagnant and Kockelman, 2015, 2014; Greenblatt and Shaheen, 2015). The fact that this technology is driverless and accordingly available every time and every place makes us assume that they are rather shared and consequently easily accessible and affordable for all (Fagnant et al., 2015; Harper et al., 2016; Meyer et al., 2017). After several years of growing interest in AVs by car manufacturers investing millions of dollars to develop driverless cars, gaining expertise of operation of future shared autonomous vehicles (SAVs) is a major concern. Their goal is to play the role of an operator of vehicle sharing systems with new business models capturing profit per kilometer or per trip (Burns et al., 2013; Firnkorn and Müller, 2012; Stocker and Shaheen, 2018). Although several SAV services are conceivable, it yet is largely unclear which ones will prevail in term of efficiency and effectiveness.

Today's mostly experimental SAV services are modified versions of ordinary electric cars with four or five seats inside. However, it is still uncertain how the most efficient design of these vehicles would look like. In addition to the vehicle characteristics and features, operational aspects of this new service could intensely influence its success. Some configurational characteristics such as fleet size, allocation and relocation strategies, service area, and infrastructures have a direct impact on the parameters that are important for mode choice decision of travelers. Although travel time and cost are the most important parameters in this regard, in the case of a shared system, some other parameters such as wait times and detour times (in the ride share mode) are of great significance. Moreover, the variation of individuals' attitude toward using this system may significantly affect service performance. In particular, the absence of a driver in SAV may generate an important concern for travelers and consequently result in lower demand. Thus, all of these parameters have to be considered in the upstream planning for having an accurate estimation of service performance measures.

Thanks to recently developed approaches, especially multi-agent simulation, parameters important for mode choice decision of travelers can be reflected at a fine-grained level. Potentially, earlier multi-agent activity-based simulations are able to consider the complex supply-demand relationships of the multi-modal transportation system. Various aspects of operating future SAVs are the subject of current research efforts based upon this approach. In particular, some in-depth investigations have been recently carried out on SAV fleet optimization, rebalancing, and cost structures of operational models (Bösch et al., 2018; Hörl et al., 2019; Loeb and Kockelman, 2019). In several studies, the human-related side of driverless cars and their impacts on service demand have been assessed (Kamel et al., 2018; Vosooghi et al., 2019). The impact of ridesharing on the operational efficiency of SAV has also been the subject of few investigations (Farhan and Chen, 2018; Hörl, 2017). However, to the best of our knowledge, none of these studies considers all affecting aspects of SAV operation at the same time. The present research addresses this gap by conducting comprehensive dynamic demand simulations in a multimodal network. The analysis of given simulation outputs allows investigating the effects of different operational components and vehicle specifications (specifically vehicle capacity) on the efficiency of the offered service, considering dynamic demand responsive to the network and the level-of-service (LoS) by integrating user taste variations and value of travel time (VTT). The further contribution of the present research is designing an SAV service upon a real-world case study and evaluating its performance using more relevant metrics. To respond to the existing needs in the simulation and analysis, some features on traveler behavior and vehicle allocation and relocation strategies are modified or developed beforehand. Simulation experiments are based upon the real data for the transportation system of the Rouen Normandie metropolitan area in France using the multi-agent transport simulation platform (MATSim).

---

[*] Corresponding author. Tel.: +33-6-6597-4453.
 *E-mail address:* reza.vosooghi@irt-systemx.fr

The remainder of this paper is structured as follows. In Section 2, a review of the relevant literature on this topic is presented. In Section 3, we present the model specification and setup process. In Section 4, overall results, as well as detailed analysis categorized by each service aspect, are presented. Finally, in Section 5, insights gained through this research are discussed and suggestions for further work are given.

## 2. Prior research

To date, numerous investigations have been conducted on SAV demand modeling and simulation particularly in the last 5 years. Several approaches have been developed to anticipate the demand for future SAV services. These approaches fall into two main categories: (i) survey and analysis, and (ii) agent-based simulation. The first approach is mainly used to produce rough estimations of potential demand based on stated preferences surveys (Bansal et al., 2016; Haboucha et al., 2017; Krueger et al., 2016). The second approach is widely employed for various study purposes and particularly those related to the supply side of SAV services in the context of dynamic demand simulation. The simulations based on the latter approach involve two different multidimensional decision processes: (i) discrete choice modeling and (ii) co-evolutionary algorithm. Discrete choice modeling is based on the assumption of random utility maximization, but co-evolutionary algorithm is more relied on finding stochastically the maximized utility for various choice sets including not only travelers' route and mode choice decisions, but also a set of activity decisions. The second group of simulations is known as activity-based multi-agent simulation. Given the purpose of this study –i.e., conducting comprehensive simulations considering all factors affecting the designing of the SAV services – we review multi-modal simulations incorporating dynamic demand that are responsive to the network and traffic.

Several studies have integrated SAVs into the area where private cars are not allowed or they are all replaced by the new service. Azevedo et al. (2016) proposed an integrated agent-based traffic simulator built on disaggregated behavior models in both demand and supply (SimMobility) to study the potential impacts of introducing autonomous mobility on demand (AMoD) service in a car-restricted zone of Singapore. In this work, individual preferences to use autonomous vehicles were kept unchanged and only the cost of the service was assumed as 40% less than the regular cab service of Singapore. The studied AMoD system employs autonomous mid-size sedans without sharing rides. Their simulation is performed through some optimization processes in terms of facility location, vehicle assignment and routing, and vehicle rebalancing. Their results suggest that rebalancing results in higher demand. In addition, the passenger waiting time is strongly correlated with the fleet size and number of parking stations. However, further growth of those variables has no more impacts once an optimal demand is reached. Heilig et al. (2017) used an agent-based travel demand model with macroscopic traffic simulation to evaluate the transportation system of the Stuttgart region where all the private cars are replaced by an AMoD service. They performed simulation for more than one day (one week) and analyzed the changes in overall transportation system performance. Furthermore, the fleet required to fulfill the demand is investigated. In their simulation, the cost per mile of a proposed service is assumed 70% less compared to the private cars and the user preferences are kept unchanged. The simulation encompasses the relocation strategy during nighttime, and it is shown that total vehicle mileage decreases up to 20% after the introduction of a new AMoD service. Martinez and Viegas (2017) tried to explore the potential outcomes of so-called radical change in urban mobility configuration of Lisbon city after introducing Shared Mobility services based upon a spatially aggregated agent-based simulation. In the simulated scenarios, all private mobility and conventional buses are replaced by Autonomous Shared Taxis and Taxi-Buses. Their simulation incorporates several optimization models in order to assign dynamically the vehicles or generate them if needed for a given day. Based on their results, it is inferred that congestion and emissions would strongly decline by introducing those shared services. Chen and Kockelman (2016) employed a multinomial logit mode choice model in an agent-based framework to asses various dynamic pricing strategies on mode shares estimate of electric SAV in Austin, Texas. Due to the spatial aggregation, the mentioned study ignores trips under one mile and non-motorized modes. Since SAV travelers can use their in-vehicle time to do other activities, the value of travel time (VTT) for this mode in the mentioned study is considered variable and dissimilar to transit. The simulation includes private cars, transit, and electric SAV. According to the performed analysis, electric SAV modal share changes significantly by variations of VTT and service fares. Besides, it is shown that some service operational metrics can be improved via targeted pricing strategies. Wen et al. (2018) in a comprehensive study investigated the deployment of AV and SAV services as the last-mile solution focusing on operation design. They employed a detailed nested logit structure for the mode choice model. In their study, an agent-based simulation is used to estimate the LoSs. They showed that there exists an important trade-off between fleet size, vehicle occupancy, and traveler experience in terms of service availability and response time. Although the mentioned research incorporates SAV user preferences by varying the alternative specific constant in the mode choice model, it includes only the unobserved (undetected) parameters of mode choice decision and neglects user specific attributes.

All the above-mentioned studies incorporate discrete choice modeling as a traveler decision choice mechanism. In some other studies, however, utility scoring is used instead. Hörl (2017) utilized MATSim to evaluate the dynamic demand response of autonomous cab service. This researcher integrated two service operators into the simulation and system performance and compared operational indicators. A fleet of 1000 AVs is introduced to the transportation system of the city of Sioux Falls in all scenarios. The simulation results reveal that the service with ridesharing attracts a larger number of travelers at off-peak hours. The user taste variation in the mentioned study is kept unchanged. In our previous work, we used the same framework (MATSim) to explore the impact of user trust and willingness-to-use on fleet sizing of SAV service integrating to the transportation system of Rouen Normandie metropolitan area (Vosooghi et al., 2019). The simulation is performed using the categorized utility scoring according to the individual sociodemographic attributes of users. Our previous work incorporates the first dynamic demand simulation of SAV service considering the user taste variation. The mentioned study shows the significant importance of traveler trust and willingness-to-use varying the SAV service use and the required fleet size. This work benefits of several optimization models to assign the vehicles dynamically. However, the study incorporates SAV services without ridesharing and rebalancing strategy.

Table 1 presents a summary of the mentioned studies stating their objectives and the main features.

**Table 1**
Summary of the selected literature on SAV demand modeling and service simulation.

| Author(s), year | Demand estimation approach/ Mode choice mechanism | Vehicle characteristics | SAV user preferences | Assessment purposes |
| --- | --- | --- | --- | --- |
| Azevedo et al., (2016) | Activity-based multi-agent simulation/Hierarchical discrete choice modeling | Mid-size sedans w/o ridesharing | Unvaried | Impact assessment, determine the fleet size and parking stations requirements |
| Chen and Kockelman, (2016) | Activity-based multi-agent simulation/Multinomial logit mode choice modeling | NM | Variable (willingness to pay, the value of travel time) | Sensitivity assessment of pricing strategies E-SAVs mode shares |
| Heilig et al., (2017) | Activity-based multi-agent simulation/Discrete choice modeling | Standard 4-seats w/ ridesharing | Unvaried | Impact assessment, determine the fleet size |
| Martinez and Viegas, (2017) | Trip-based multi-agent simulation/Discrete choice modeling | 6-seats minivan w/ ridesharing | Variable (car ownership, public transport pass) | Impact assessment, impacts on car fleet size, the volume of travel and parking requirements, $CO_2$ emissions |
| Hörl, (2017) | Activity-based multi-agent simulation/Utility scoring | Standard 4-seats w/ and w/o ridesharing | Unvaried | Dynamic demand response simulation of AVs and SAVs |
| Wen et al., (2018) | Trip-based multi-agent simulation/Nested logit mode choice modeling | Standard 4-seats w/ and w/o ridesharing | Variable (intrinsic preference) | Design of last-mile AV and SAV services integrated to public transit |
| Vosooghi et al., (2019) | Activity-based multi-agent simulation/Categorized utility scoring | Standard 4-seats w/o ridesharing | Variable (age, gender, and household income) | Impact assessment of user preferences on individual-ride SAV fleet sizing |

Some other studies investigate the use of SAV services in a multi-modal system incorporating various dispatching strategies or pricing schemes. However, the demand of the proposed services in these simulations is not necessarily dynamic (Auld et al., 2017; Farhan and Chen, 2018; Hörl et al., 2019) or responsive to the traffic states (Chen et al., 2016; Fagnant and Kockelman, 2018, 2014). Many other studies incorporate static or predefined demands (Boesch et al., 2016; Fagnant and Kockelman, 2014; Levin et al., 2017) or simulate only one mode (Bischoff and Maciejewski, 2016; Loeb and Kockelman, 2019; Zhang et al., 2015a). There are also large in-depth investigations on AV dynamic assignment (Hyland and Mahmassani, 2018), ride-share matching optimization problem (Agatz et al., 2011; Alonso-Mora et al., 2017), and SAV rebalancing and ridesharing (Spieser et al., 2014; Zhang et al., 2015b). These studies focused rather on optimization problems and ignored mode choice mechanism in a multi-modal context or time-dependency in travel time caused by congestion.

As shown in Table 1, most of earlier comprehensive simulations to investigate the operation of SAV service are based on the homogeneous structure of behavior in terms of sociodemographic attributes, except our previous work (Vosooghi et al., 2019) that incorporated only individual ride service. Hence, the vehicle characteristics and specifically vehicle capacity and its impacts on SAV service performance have remained a missing component in all prior studies. Service cost and the need for enabling rebalancing strategy have similarly received low attention. Considering all mentioned SAV simulation features, to the best of our knowledge, the present study is the first comprehensive investigation of a real-world scenario that could provide new insight into the design of such service.

## 3. Model specification and set up

### 3.1. Simulation framework

In this work, the multi-agent transport simulation (MATSim) (Horni et al., 2016) and its Dynamic Vehicle Routing Problem (DVRP) extension (including the Demand Responsive Transport (DRT) application) (Bischoff et al., 2017; Maciejewski et al., 2017) are used. The main idea behind MATSim is the simulation of an artificial population, represented by agents, who perform their respective plans including activities and movements between activity locations throughout a day. The movements are simulated within a dynamic queue-based model in which all agents interact dynamically with each other in a network (traffic simulation). At the end of the simulation day, which usually exceeds 24 hours due to the longest activity chain, all agents evaluate the performance of the executed plan by measuring and scoring the deviations from the initial plan and the utility of using a mode. This process is called "utility scoring". In the next iterations, agents try to maximize their scores by modifying their plans. This "re-planning" process is performed rather by using another mode and route or by ending an activity sooner than at its planned end time. The more agents explore potential alternatives, the more they learn about their optimal plans. Once the convergence on the total score is reached, agents stop to innovate their plans and try to select one plan from their memorized set of plans and to find out the plan with the best score. This is repeated until a systematic relaxation is reached.

Re-planning is usually done after the plan execution and traffic simulation. However, for the simulation of new transportation systems and specifically those that need a cyclic re-computation of vehicle tasks and routes (e.g. SAV, on-demand services with multiple requests), instant decisions must be made while the traffic simulation is running. Such a decision making is possible is possible using dynamic agent module included in DVRP-DRT extension (Maciejewski, 2016), which directly interacts with the traffic simulation of MATSim.

### 3.2. Ridesharing and rebalancing

In the present study, we used the dispatch algorithm of ridesharing developed by Bischoff et al. (2017) that performs a centralized on-the-fly assignment of vehicles to on-demand requests. This optimizer returns a list of requests and vehicle paths between pick-up and drop-off points. In order to route SAVs dynamically, an insertion heuristic that aims to minimize the total SAVs workload is employed. The SAVs workload is measured as the total time spent on handling requests. This leads to a lower detour for each user. The optimization process seeks also to decrease vehicle usage for more requests, which results in more service availability and consequently greater demand. During the simulation, when a new request is submitted, the algorithm searches the routes of all vehicles for optimal insertions. An insertion is feasible when it satisfies the following conditions: (i) the overall travel time constraints (including waiting and in-vehicle times) are satisfied for already inserted requests (passenger(s) on board) and (ii) the expected boarding times for the awaiting and upcoming requests need to remain within a defined time frame. All feasible insertions are then evaluated and the first insertion that offers the smallest increase of vehicle work time will be selected. If no feasible insertion is found, the request is rejected. A request can be rejected (e.g., due to constraints violation) only immediately after submission, and already accepted requests cannot be rejected or re-scheduled.

We employed the rebalancing strategy that is included in the DRT extension of MATSim, which is based on the Minimum Cost Flows problem. In this problem, one seeks to "optimize" the time-varying flows on each arc between aggregated demand hubs and idle vehicles, taking into account congestion effects along arcs and at nodes. Idle vehicles are relocated in regular intervals according to the estimated demand of the previous iteration. The expected demand for the next 60 minutes is considered in the optimization process.

It is noteworthy that the selected dispatch and rebalancing algorithm may have a strong impact on service performance indicators (Hörl et al., 2019; Hyland and Mahmassani, 2018). Particularly, when the demand for SAV service is relatively high, applying simplified assignments (e.g. FIFO) can lead to the worst service efficiency (Hyland and Mahmassani, 2018). However, we found that the employed strategies for vehicle assignment and relocation are accurate enough for our purpose. Furthermore, it is of note that the multi-agent simulation already adopts heuristic rules in feedback loops to achieve approximate convergence and consistency between multidimensional decisions and network loading. Thus, it may require even more computational resources to achieve equilibrium when very sophisticated heuristic rules are applied to find good assignment and relocation of vehicles.

*3.3. Inputs and model setup*

As mentioned earlier, the main goal of the present study is to design an SAV service considering all affecting operational and user-related aspects. For this purpose, the simulation inputs are based on real activity chains replicating the traveler patterns and schedules derived from the transport survey and census. Fig. 1 illustrates the overall framework. A synthetic population for the case study area is generated using an open source generator developed in our work (Kamel et al., 2018) that applies fitness-based synthesizing with multilevel controls. Some major attributes such as age, gender, household income range, and socio-professional category are used for controls. These are the attributes with an important impact on SAV mode choice (Al-Maghraoui, 2019) or are the joint attributes of synthetic population and activity models. The activity chains are then allocated to each synthetic individual according to their socio-professional attributes. Based on transport survey analysis conducted for two French case study areas (Paris and Rouen Metropolitan area), we found that the activity chains are significantly correlated to those attributes (Kamel et al., 2019; Vosooghi et al., 2019). The socio-professional category consists of six groups of persons: employed, unemployed, students, people under 14 years, retired, and homemakers. The generated synthetic population is validated by comparing relative errors of the synthetic and real population in each zone for estimated and given marginal data of each attribute.

For two main trip purposes (work and study), the fine-grained geographical zones of activity are given in the census data. For other trip purposes, an origin-destination matrix based on the transport survey is estimated. Both of these data are employed in the process of activity chain allocation. In the latter case, for each trip origin in each zone, a destination zone according to the probability of trip purpose by socio-professional category is allocated. Then, the activity's precise locations are randomly appointed along the zone in keeping with existing activity types and land-use category.

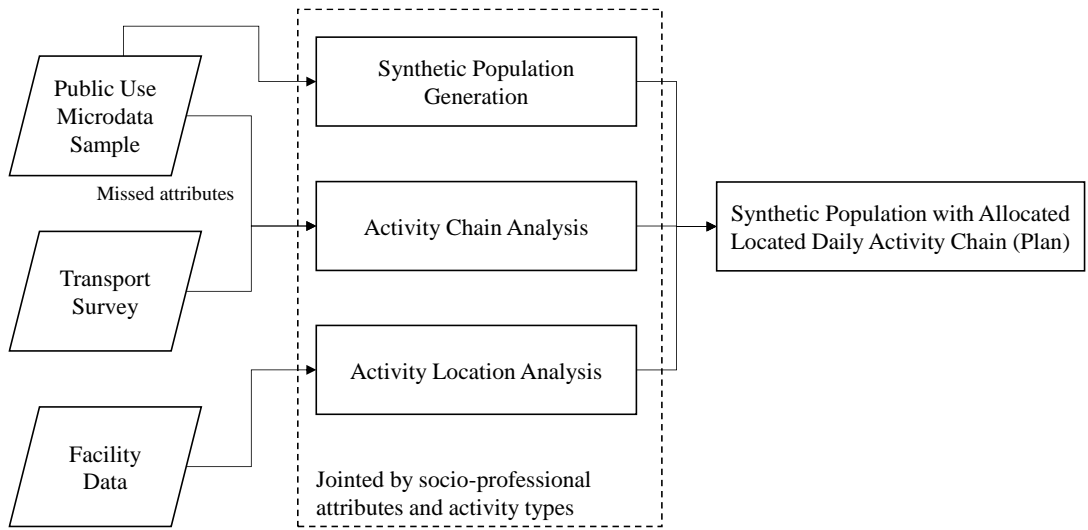

**Fig. 1.** Synthetic population generation and activity chaining framework.

In MATSim, utility scoring is performed based on the Charypar-Nagel scoring method (Charypar and Nagel, 2005). The function includes both activity and leg scores. In conducted simulations, due to the lack of data only legs' scoring utilities are set according to the utility functions estimated from the recent local transport survey (EMD Métropole Rouen Normandie 2017) by employing a logit model. In addition to travel time (including waiting time) and travel cost, user's car ownership, as well as parking availability at destination, were found to be significant parameters in the mode choice model. Thus, they are incorporated into utility scoring. The detailed list of parameters as well as the estimations is available in Vosooghi et al. (2019).

In the ridesharing algorithm, we set up the detours considering that the ride times can be extended up to 30% of the direct distance. Bigger detour times for passengers are allowed only if their waiting times do not surpass 15 min. However, in that case, the SAV ride is more penalized in terms of utility (scoring). During the simulation, the idle SAVs are rebalanced every 5 min. The size of square cells used for demand aggregation is assumed to be 1 km.

*3.4. User taste variation*

User taste variations were integrated into the model based on the methodology proposed in our previous work (Vosooghi et al., 2019). In order to set up the model according to the travelers' perception and a tendency toward using SAVs, an online survey was made in (Al-Maghraoui, 2019). Users' trust and subjective criteria behind their willingness-to-use were identified. We found that the socio-professional and three other sociodemographic attributes (i.e., income, age and gender) are significant to SAV taste variations among individuals. For instance, the above-mentioned survey shows that in general men are more likely to use an SAV than women. Similarly, younger persons are more likely to use SAVs compared to older ones. In order to integrate these variations into the simulation, all the utility scoring and functions are estimated and set up separately for each group of users within the same socio-professional category. The marginal time and cost-varying parameters in the scoring function are then multiplied by the factors of user trust and willingness-to-use so that the score (utility) of SAV for the similar trips varies according to the sociodemographic attributes of travelers including age, gender and household income range.

*3.5. Simulated scenarios*

In order to apply our simulation to design an SAV service for a real-world scenario, we chose to consider the Rouen Normandie metropolitan area (France) as the case study. Rouen was a fitting venue for at least three reasons. First, we had access to the most recent transport survey (EMD Métropole Rouen Normandie 2017). Second, the population (about 500,000 inhabitants) and the metropolitan area network sizes allow us to perform the simulation with an acceptable downscaling rate (10%), which results in quite accurate outputs compared to the full-scale model (Bischoff and Maciejewski, 2016). Actually, in some studies relying on agent-based simulation and utility scoring, the population of case study areas is highly downscaled (1%) due to the high computational time (Hörl et al., 2019; Kamel et al., 2019). This extensive downscaling may potentially affect the service performance evaluations considering the spatiotemporal interaction of supply and demand in large study areas. The third reason for choosing this area is that some experiments on self-driving cars are currently undertaken; thus, it is possible to gain data on traveler behavior in a near future and to integrate them to the extended simulations. Furthermore, Rouen Normandie metropolitan area is a promising candidate for replacing existing private modes with an SAV service, especially in the Rouen old town.

To support the simulation of such a scenario, the synthetic population from the Public Use Microdata Sample (INSEE 2014) was generated. Based on the local survey including 5,059 households and 11,107 individuals, 929 activity chains including eight trip purposes were found. As mentioned before, the activity chains and time profiles were allocated to the synthetic individuals according to their socio-professional category.

The simulations were run for several fleet-size and fleet-capacity scenarios with and without considering ridesharing or rebalancing strategies to appreciate system performance metrics. Regarding SAV, prices of 0.5 Euro per kilometer for the individual ride and 0.4 Euro per kilometer (direct distance) for ridesharing services are assumed. These service prices are slightly more expensive than private car ride costs in France (0.3 Euro per kilometer - DG Trésor (2018)). Moreover, they are almost similar to the values that have been estimated or concluded in other investigations. For instance, Chen et al. (2016) estimated the price for electric SAV from $0.42 to $0.49 per person-trip-mile and Bösch et al. (2018) estimated it 0.43 CHF per passenger kilometer. In our study, the SAV service waiting time is integrated to the utility scoring and the value of waiting time is considered 1.5 times larger than the value of in-vehicle travel time (Wardman et al., 2016). The SAV fleet is initially distributed from four fixed points inside the city and out of old town. Therefore, no "warm-start" - as in Fagnant and Kockelman (2014) - or random distribution are considered.

*3.6. Performance metrics*

Although a limited number of studies have simulated SAV service incorporating its dynamic demand, there are several investigations on the performance evaluation of such a system. In this regard, a long list of performance metrics has been used as well. These metrics do not necessarily have the same consequences. For instance, Fagnant and Kockelman (2014) used traveler wait times in order to estimate required fleet sizes to serve various predefined demands. Since the demand in their study is considered static, the wait times could be a relevant indicator to evaluate the service performance. However, as shown in a more recent work of the same authors, the lower in-vehicle and wait times in a simulation enabling dynamic ridesharing result in higher excess vehicle kilometer traveled (VKT) (Fagnant and Kockelman, 2018) and therefore cannot be the only relevant indicators for the fleet sizing. In this regard, traveler wait times have been used as a key indicator to define the optimum scenario in some other dynamic demand simulations (e.g. in Azevedo et al. (2016) or Chen and Kockelman (2016)). However, in the current research we use the term "fleet in-service rate" for the fleet sizing. This indicator is defined as the number of occupied or in-service vehicles (including going to pick up a client) over all vehicles. The other metric representing the proportion of extra VKT (due to the unoccupied or rebalancing mileage) over total VKT will be used in parallel to evaluate empty vehicle traveling distances and to find the balanced trade-off between these two indicators. The detour distance would be among the main traveler-related indicators representing extra travel distances due to the shared rides. The traveler wait times here will be used as the LoS evaluation; the lower the wait time, the higher the service level is. In fact, due to the dynamic decision mechanism between available alternatives for each traveler, higher wait times result in lower SAV demand and consequently service usage. Therefore, this parameter implicitly affects the main performance indicators. With the aim of comparing the service revenues for different scenarios, the in-vehicle passenger kilometer traveled (PKT) is defined. This indicator presents the sum of trip distances traveled by each individual on SAVs. In order to investigate the usage pattern of SAV service in the case of ridesharing, the "on-board occupancy rates by a number of passengers" is proposed. Other metrics used in the dynamic demand simulations to evaluate the performance of proposed service are the number of persons or vehicle trips (Chen and Kockelman, 2016; Heilig et al., 2017) and average in-vehicle times (Hörl, 2017; Martinez and Viegas, 2017). However, these performance indicators are descriptive rather than consequential and thus will not be used for the fleet sizing and vehicle capacity determination.

To evaluate the overall performance of the transportation system, mode share indicator as in the majority of other studies will be compared. Although this research does not incorporate environmental impact measurements, the total distances driven by car and SAV are estimated and compared for all scenarios to illustrate how the congestion would change after the introduction of different SAV services. It is of note that due to the high uncertainty of future SAV service and infrastructure costs, in the present study only transport-related indicators are evaluated and analyzed.

## 4. Case study results

### 4.1. Overview

A base-case scenario (S0) simulation run was conducted without integrating SAV and calibrated using the actual modal shares of the case study area. Next, the SAV service was simulated for various fleet sizes (2.0 k to 6.0 k) with individual rides (S1) and for those with ridesharing strategy. In the case of ridesharing, three different vehicle capacities were suggested for the simulation: a small car with two seats (S2), standard 4-seats car (S3), and 6-seats minivan (S4). Table 2 illustrates the modal splits for all scenarios. It is noteworthy that given the low modal share of the bike (less than 0.1%) and related changes, this mode was not simulated.

**Table 2**
Modal splits estimated for all scenarios and fleet sizes[*].

| Scenario | Fleet size Mode | 2000 | 2500 | 3000 | 3500 | 4000 | 4500 | 5000 | 5500 | 6000 |
|---|---|---|---|---|---|---|---|---|---|---|
| S1- non-ridesharing | | | | | | | | | | |
| | Car | 59.3 | 58.8 | 58.5 | 58.3 | 58.0 | 57.7 | 57.6 | 57.4 | 57.5 |
| | Walk | 28.3 | 28.3 | 28.2 | 28.2 | 28.2 | 28.2 | 28.2 | 28.1 | 28.1 |
| | SAV | 3.1 | 4.4 | 5.3 | 6.0 | 6.5 | 6.9 | 7.2 | 7.5 | 7.6 |
| | PT | 9.2 | 8.4 | 8.0 | 7.6 | 7.3 | 7.1 | 7.1 | 6.9 | 6.8 |
| S2- ridesharing (2-seats small car) | | | | | | | | | | |
| | Car | 59.1 | 58.8 | 58.5 | 58.3 | 58.1 | 57.8 | 57.7 | 57.8 | 57.7 |
| | Walk | 28.3 | 28.3 | 28.3 | 28.2 | 28.2 | 28.3 | 28.3 | 28.2 | 28.2 |
| | SAV | 3.8 | 4.6 | 5.2 | 5.9 | 6.3 | 6.5 | 6.7 | 6.9 | 7.0 |
| | PT | 8.8 | 8.3 | 8.0 | 7.6 | 7.5 | 7.3 | 7.2 | 7.1 | 7.1 |
| S3- ridesharing (standard 4-seats car) | | | | | | | | | | |
| | Car | 58.9 | 58.7 | 58.3 | 58.1 | 58.0 | 57.9 | 57.8 | 57.7 | 57.7 |
| | Walk | 28.3 | 28.3 | 28.3 | 28.3 | 28.3 | 28.3 | 28.3 | 28.3 | 28.3 |
| | SAV | 4.0 | 4.6 | 5.3 | 5.9 | 6.0 | 6.4 | 6.6 | 6.8 | 6.8 |
| | PT | 8.7 | 8.3 | 8.0 | 7.7 | 7.6 | 7.4 | 7.3 | 7.2 | 7.2 |
| S4- ridesharing (6-seats minivan) | | | | | | | | | | |
| | Car | 59.1 | 58.8 | 58.4 | 58.2 | 58.0 | 57.9 | 57.8 | 57.8 | 57.7 |
| | Walk | 28.2 | 28.3 | 28.3 | 28.3 | 28.3 | 28.3 | 28.3 | 28.2 | 28.2 |
| | SAV | 4.1 | 4.6 | 5.4 | 5.9 | 6.1 | 6.4 | 6.8 | 6.7 | 6.9 |
| | PT | 8.6 | 8.3 | 7.9 | 7.6 | 7.5 | 7.4 | 7.3 | 7.2 | 7.1 |

[*] Due to the rounding process for each modal share, the sum could exceed or be less than 100%.

As shown in Table 2, the modal shifts toward an SAV service come from both public transport and car modes. This shift is consistent with findings in the literature (Chen and Kockelman, 2016; Hörl, 2017; Martinez and Viegas, 2017; Wen et al., 2018). However, in the case of big SAV fleet sizes, the public transport mode share decreases significantly relative to the car. This reduction is due to the utility of the proposed service, which is rather similar to the public transport mode. The service cost is also an important factor that encourages public transport users to choose a service that costs a bit more but is more appealing due to the lower travel time. Table 2 illustrates an interesting result regarding SAV modal share evolution. As can be expected, by increasing the fleet size, SAV modal shares increase accordingly. However, this growth does not follow the same trend for all scenarios. While SAV modal share in scenario 1 (individual ride) is the lowest one among all scenarios in the case of the smallest fleet size, this metric is conversely the highest for the fleet size of 6.0se k vehicles. This result can be explained by the presence of a balanced trade-off between service cost, demand (which affects waiting time) and extra in-vehicle time due to detour distances. In fact, when the waiting time is more important compared to in-vehicle times, which is the case for small fleet sizes, the time-based cost of service could surpass the service cost for users. Therefore, the SAV demand and consequently its modal share decreases. However, in the case of big fleet sizes, as the waiting time is not as important as in the case of small fleet sizes, the in-vehicle time (including detour time) becomes an important factor for the decision-making.

Table 3 presents the evolution of total driven distance including private cars and SAVs. By deploying SAV services, this indicator increases in all scenarios. Clearly, having a bigger fleet size in each scenario results in more use of vehicles. However, scenarios with ridesharing strategy have lower total driven distance compared to individual ride (except for the scenario with 2.0 k SAVs). This difference is attributed to the higher SAVs' occupancy rates in ridesharing scenarios. Comparison of vehicle capacities shows that in the scenarios with the fleets of 4-seats and 6-seats SAVs, the total driven distance is slightly lower than when 2-seats small SAVs are simulated (except for the scenario with 2.0 k SAVs). Meanwhile, for some fleet sizes, this indicator has the lowest value when 4-seats standard cars are used. The shorter total driven distance of individual ride service and smaller vehicles in ridesharing scenarios when 2.0 k SAVs are simulated is because of the relatively much lower service demand, which is due to the low LoS provided in that fleet size.

**Table 3**
Total driven distance including car and SAV modes (million kilometers).

| Fleet size | S0- base-case scenario | S1- non-ridesharing | S2- ridesharing (2-seats small car) | S3- ridesharing (standard 4-seats car) | S4- ridesharing (6-seats minivan) |
|---|---|---|---|---|---|
| - | 8.88 | - | - | - | - |
| 2000 | - | 10.05 | 10.06 | 10.14 | 10.17 |
| 2500 | - | 10.50 | 10.33 | 10.31 | 10.30 |
| 3000 | - | 10.77 | 10.43 | 10.43 | 10.45 |
| 3500 | - | 10.95 | 10.61 | 10.50 | 10.53 |
| 4000 | - | 11.12 | 10.62 | 10.48 | 10.53 |
| 4500 | - | 11.20 | 10.66 | 10.55 | 10.53 |
| 5000 | - | 11.26 | 10.60 | 10.53 | 10.56 |
| 5500 | - | 11.31 | 10.73 | 10.60 | 10.59 |
| 6000 | - | 11.32 | 10.69 | 10.58 | 10.63 |

Table 4 presents a summary of the average number of rides per SAV for all scenarios. As can be seen, in the case of small fleet sizes, the SAVs with bigger capacity satisfy more requests because the expected waiting time is relatively high enough to play a major role in the mechanism of SAV mode choice decision. In the scenarios of SAVs with ridesharing and bigger vehicle capacity, the expected waiting time is shorter compared to the scenarios of the individual ride, thus more rides are satisfied. However, by increasing the number of vehicles, the detour distance becomes a more important parameter and therefore in the scenarios with more places in the vehicles, slightly fewer requests are observed. It should be noted that the expected waiting and detour times are the parameters that are estimated for each synthetic individual (agent) during the simulation. Agents learn about their plans (including trips and activities) and final decisions are made in the last iteration when convergence is reached. Thus, the real and expected waiting and detour times have different values. Since the number of expected values is huge and we cannot visualize them all, we present only real values in this paper.

**Table 4**
Average number of rides per SAV.

| Fleet size | S1- non-ridesharing | S2- ridesharing (2-seats small car) | S3- ridesharing (standard 4-seats car) | S4- ridesharing (6-seats minivan) |
|---|---|---|---|---|
| 2000 | 11 | 16 | 17 | 17 |
| 2500 | 16 | 17 | 17 | 17 |
| 3000 | 17 | 17 | 18 | 18 |
| 3500 | 17 | 17 | 17 | 17 |
| 4000 | 17 | 16 | 16 | 16 |
| 4500 | 17 | 16 | 15 | 15 |
| 5000 | 16 | 14 | 14 | 14 |
| 5500 | 15 | 14 | 13 | 13 |
| 6000 | 14 | 13 | 13 | 13 |

As shown in Fig. 2, the average detour time varies between 4 and 7 minutes in all scenarios. Likewise, the average in-vehicle time varies from 37 to 48 minutes. However, the variation of waiting time (excluding the smallest fleet size) for each fleet size remains very slight. It is noteworthy that since the simulations are dynamic-demand, low average waiting times for small fleet sizes are due to the low SAV demand especially during peak hours (particularly in S1 and S2). In fact, for fleet size below a certain size, the expected waiting time increases considerably. Therefore, SAV mode becomes less competitive to other available alternatives in terms of generalized cost, except in the morning peak hour when the LoS of other alternatives are as low as SAV (Fig. 3). Thus, the demand for SAV service and consequently estimated waiting time decreases. By increasing the number of vehicles, the expected waiting time declines. However, this time is shorter than a critical waiting time (a value that makes SAV non-competitive in terms of utility), its impact on SAV mode choice becomes minor. As a result, the estimated waiting time follows a very slight decreasing trend particularly in ridesharing scenarios. In the non-ridesharing scenario, the estimated average waiting time falls faster for big fleet sizes. This faster decline by increasing the number of vehicles is explained by the fact that for each request an available SAV can be found in lower direct access distance; however, in the ridesharing scenario, this SAV may not necessarily be without passenger and therefore a relatively higher waiting time is required. Thus, the decrease in average waiting time by enlarging fleet size in ridesharing scenarios becomes slighter.

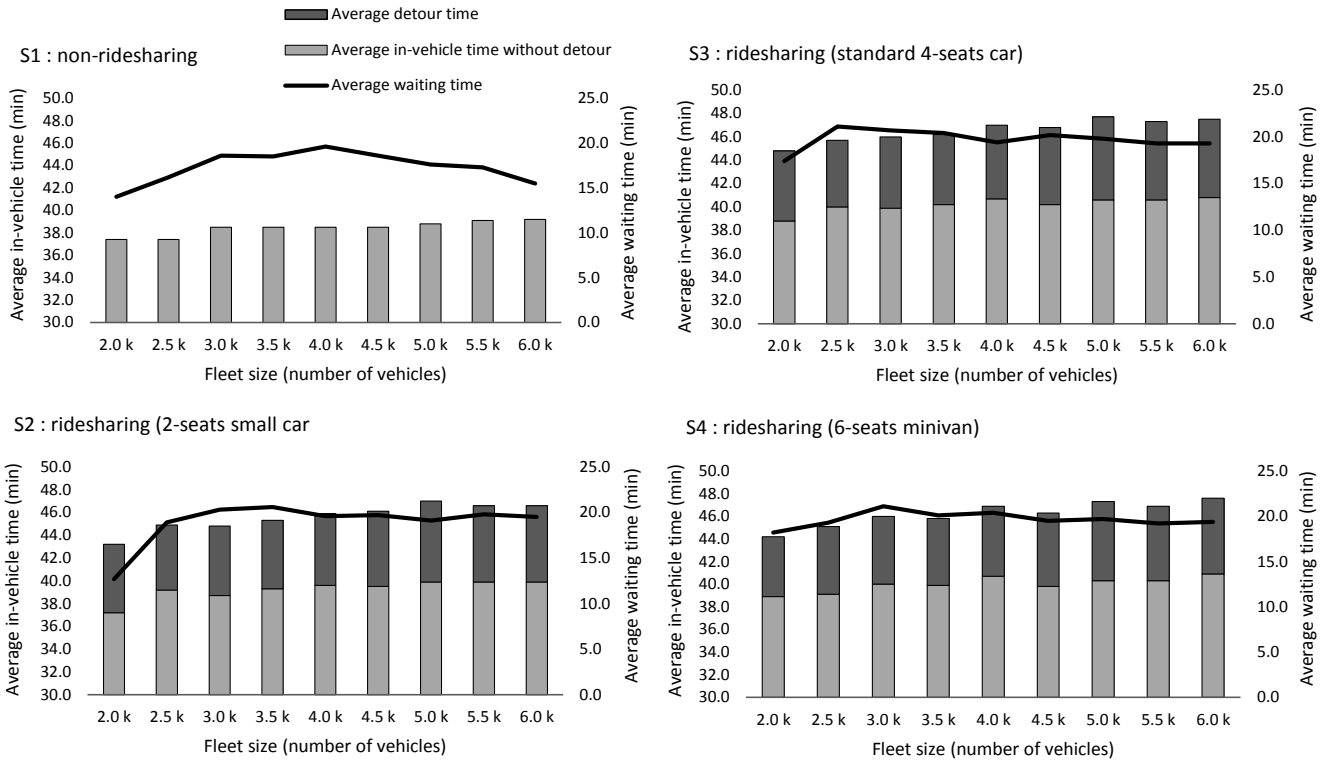

**Fig. 2.** SAV service time variables for all scenarios and fleet sizes.

*4.2. Fleet size*

The bigger fleet size and accordingly the higher SAV modal share do not necessarily lead to a better-optimized operation. In fact, a trade-off between overall expenses and revenues has to be balanced. Service costs include capital expenditure (CAPEX) and operating expenses (OPEX). Since this research does not incorporate infrastructures of SAV service, CAPEX is assumed directly correlated with the fleet size. OPEX is however associated with fleet usages and mileage. Fig. 3 illustrates the hourly fleet in-service rates for all scenarios and various fleet sizes. As shown in this figure, consistent with daily trip patterns, two peak service usages occur for morning and evening peak hours. However, unlike the SAV modal share, the service use is decreased for big fleet sizes. In fact, by increasing the fleet size, once the fleet usage becomes no longer saturated in the morning peak hour, the latter decreases quickly. This occurs by improving the LoS indicators (waiting time or accessibility in this case) and leads to the demand growth. However, this demand is somehow limited to the number of people who are already likely to choose this service compared to other alternatives that they have, even if the waiting time is very low. Similarly, if the fleet size is small and the LoS is accordingly low, users try to find a more appropriated mode. As a result, as shown in Fig. 3, for the small fleet sizes and especially in scenario 1 (individual ride), the fleet usage decreases abruptly. Again, we emphasize that the simulation results present the indicators when the interaction of service demand and supply is iteratively relaxed. In other words, the agents have already experienced the SAV service for various level of demands. Agents also tried to slightly modify their activity end time and to depart sooner in order to arrive to the next activity on time. However, the memorized expected waiting time is supposed to be high for many travelers specifically in the case of small fleet sizes. Consequently, SAV is not as used as in the case of medium and large fleet sizes.

As presented in Fig. 3, in individual ride scenario (S1), SAV service reaches the maximum fleet usage at least for one hour in some fleet sizes. This maximum use, however, does not occur for ridesharing scenarios due to two main reasons. First, in ridesharing scenarios in peak hours there are always SAVs with available seats in the acceptable distance for any requests. Second, since there is no rebalancing strategy in those scenarios (SAVs stay at the same place where the last passenger is dropped off), some SAVs that dropped off a passenger(s) far from demand hubs remain in idle mode at that location and thus the fleet usage does not reach the maximum value. This shows the importance of enabling rebalancing strategy especially when rides are shared among travelers.

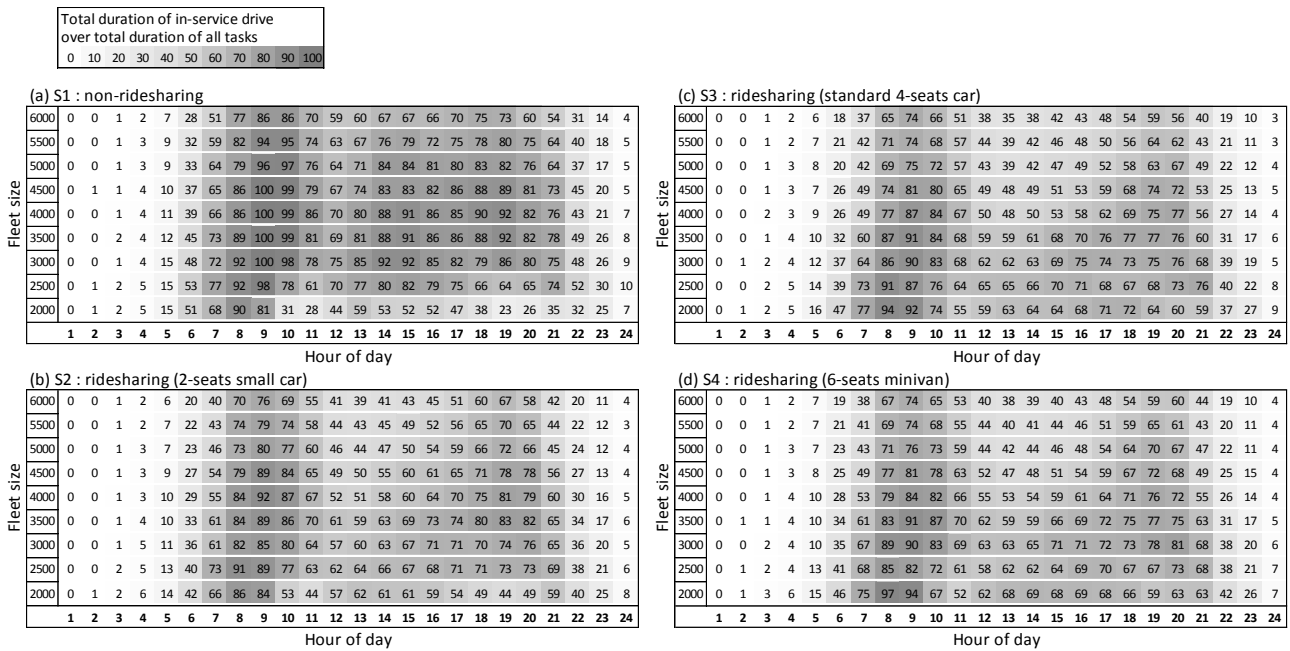

**Fig. 3.** SAV fleet hourly in-service rates.

Empty traveling distance is also a part of fleet usage. High-performance fleet size is characterized by a greater use and a lower empty distance. Fig. 4 compares average daily in-vehicle service rates (fleet usage ratio) and empty distance ratio (empty VKT over total VKT) for all scenarios and fleet sizes. As can be seen from this figure, the fleet usage fluctuates more than the empty distance ratio. In fact, by increasing fleet sizes, the empty distance ratio changes only by a maximum of 3%, meanwhile, the fleet usage drops dramatically (up to 16%). This abrupt decline may occur because there is no rebalancing strategy incorporated in those scenarios and the pick-up ride distances remain approximately within the same range of values (with lower usage and consequently more available vehicles, the pick-up ride distance becomes slightly shorter). In order to identify the best performing fleet size, two aforementioned indicators are used. Actually, regardless of any estimations about service operational cost and benefits, the fleet usage indicator can be used as the measure of effectiveness concerning CAPEX. Similarly, the empty distance ratio may be sufficiently indicative for the changes on OPEX term. Since the latter indicator stays rather constant for all fleet sizes, the best fleet sizes are identified according to the fleet usage ratio. For individual ride service, the fleet of 3.5 k SAVs, in the case of small cars with two seats, 2.5 k vehicles and for the other scenarios, approximately 3.0 k vehicles seem to be the best performing size of the fleet.

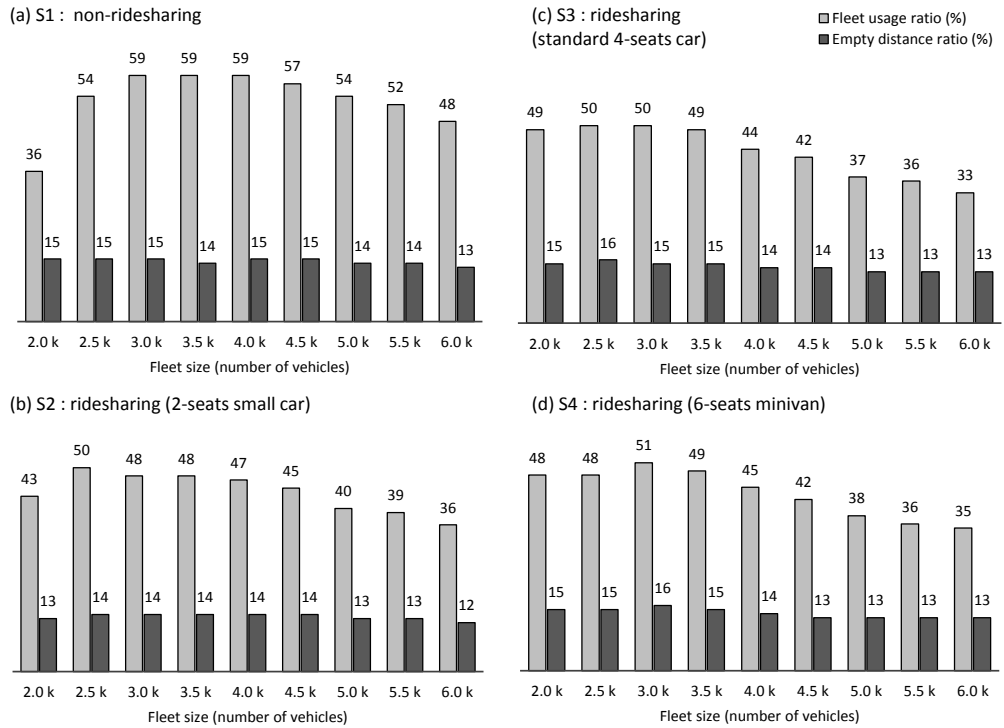

**Fig. 4.** Comparison of fleet usage and empty distance ratios for different scenarios and fleet sizes.

Another noteworthy point in Fig. 3 is the shift of the morning peak-hour service usage from 7-9 a.m. for small fleet sizes to 8-10 a.m. for big fleet sizes. This shift is actually due to two main reasons. First, in small and big fleet sizes the SAV users are dissimilar in terms of the socio-professional category to which they belong. These users have a different trip pattern, consequently by varying fleet sizes the hourly usage of SAV service changes. Second, the possibility of small changes in the activity end-time in the simulation allowed some users to leave the previous activity slightly sooner in order to arrive to the next activity with lower delay. This happens when departing earlier from an origin activity such as "Home" has not an important impact on the score of agent's whole day plan. However, when the activity at origin is "Work" or "Study", shorten those activity durations results in a much lower score and penalizes the use of SAV service. The possibility of slight changes on activity end time is enabled for all modes in the simulation, however given the score of performing activities, these changes may not surpass several minutes. Regarding different group of users, in the simulation, agents with different socio-professional category have different utility scoring. In other words, for instance, for some groups of people, the marginal utility of traveling or the value of travel time (VoT) is bigger than for other groups. Furthermore, user taste variation among the different category of travelers affects SAV mode choice. As a result, by increasing fleet sizes and improving the LoS (travel time, including wait time), the SAV service for some users with different socio-professional categories and accordingly different trip purposes become more attractive (in terms of utility). Since each activity at the destination has a dissimilar model of start-time and duration, hourly usage pattern of SAV service changes when considering a different group of users. Fig. 5 illustrates the evolution of SAV service users by their socio-professional categories for different fleet sizes. As can be seen, the ratio of employed people in all scenarios decreases by increasing fleet sizes. For students and people under 14 years of age, a slight increase in big fleet sizes is observed. Meanwhile, the changes in the ratio of unemployed people remain minor. However, there is a relatively significant growth in the use of SAV service by retired people and homemakers when the fleet size is large (especially in the case of the individual ride). By comparing scenarios of each fleet size, it can be seen that the SAV service with the individual ride is less attractive for employed and unemployed users. This occurs since the cost-related parameters of mode choice decision are more important than time-related parameters for those groups of users. The above analysis shows the importance of considering users' profile in estimating fleet hourly usage, which can potentially affect the fleet sizing.

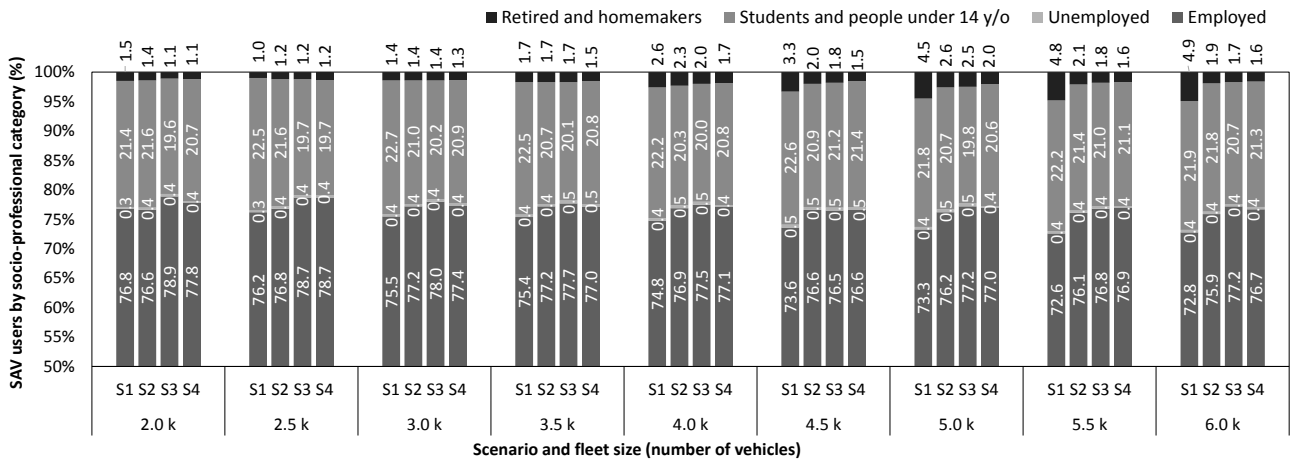

**Fig. 5.** SAV service users by socio-professional category.

### 4.3. The sharing strategy of ride

As shown in Fig. 3, in the scenarios with ridesharing, the fleets are never 100% in-service for 1-h time slices. This may occur when rides are shared but empty vehicles are not rebalanced. When a request is registered, the nearest occupied vehicle with an available seat and acceptable detour time is assigned. As a result, there are always some vehicles quite far from the demand hubs that are not consequently used for a while. In fact, ridesharing results in relatively lower fleet usage for almost all fleet sizes except for the fleet size of 2.0 k (Fig. 4). The more rides are shared, the less the fleet is used. However, each user pays for the provided services and traveled kilometers. In this case, the indicator of in-vehicle passenger kilometer traveled (PKT) may be more relevant. This indicator presents the sum of distances traveled by each individual being on-board SAVs. Fig. 6 compares SAV overall PKT for all scenarios and fleet sizes. As shown in this figure, the overall PKT of the individual ride scenario is minimum for all fleet sizes. This indicator is however almost the same for all ridesharing scenarios in the case of medium fleet sizes (i.e., 3.5 k to 4.5 k). By increasing the number of vehicles, the relative difference of PKT between individual ride and ridesharing scenarios decreases. This decline can be attributed to the high LoS provided in the case of large fleet sizes. In fact, the potential requests for the SAV service are limited. Thus, when the fleet is accessible enough for a major part of potential users, the greater service availability (occurring when the rides are shared) does not necessarily result in an important increase in demand and PKT accordingly. As a result, the growth of PKT and its differences between individual ride and ridesharing scenarios decline.

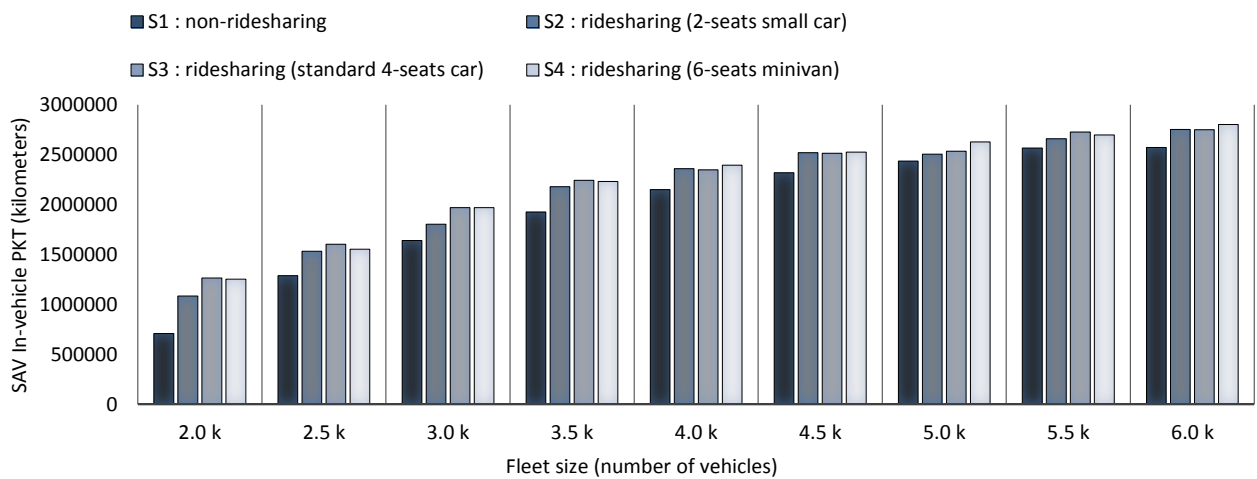

**Fig. 6.** Comparison of SAV overall PKT for all scenarios and fleet sizes.

### 4.4. Vehicle capacity

As mentioned earlier, the best performing fleet size in the case of ridesharing scenarios according to the fleet usage and empty ratios is between 2.5 k and 3.5 k vehicles. Comparing the PKT for those fleet sizes reveals that the fleet of the standard 4-seats car may be the best performing option for ridesharing. As shown in Fig. 6, for small fleet sizes, the overall PKT of the standard 4-seats car is greater. In fact, for those fleet sizes, the fleet usage is almost saturated during peak hours (Fig. 3). As a result, services are less accessible especially when the vehicle is smaller and the number of available seats is lower. It seems that the service with bigger capacity vehicles would be more used by travelers in that case; however, due to extra detour time (expected), the PKT of 6-seats SAVs is slightly less than 4-seats SAVs. In other words, for the same SAV service price, users prefer to choose a medium capacity car that has relatively shorter waiting and in-vehicle times compared to a 6-seats minivan. By increasing fleet size, as there is enough SAVs to satisfy the demand, the differences between PKTs for those scenarios become relatively minor. However, since in that case more demand is satisfied, the probability of pooling rides with an acceptable detour time becomes higher. Thus, a limited number of vehicles handle many requests in high demand areas. Meanwhile, the idle vehicles that have already dropped off a passenger far from the demand hubs stay at the same place for a while. This non-homogeneous spatial distribution of idle and high workload SAVs results in a different PKT for ridesharing scenarios with big fleet sizes. This difference occurs when the rebalancing strategy is not enabled. It is of note that in the small and medium fleet sizes, the fleet usage is relatively high and thus, SAVs are somehow rebalanced across the high demand hubs and dispersed better within the region. This shows again the importance of considering rebalancing strategy.

In order to explore the use of vehicle capacities, on-board occupancy rates by the number of passengers (PAX occupancy ratio) are compared for ridesharing scenarios. As shown in Fig. 7, for all ridesharing scenarios, by increasing the fleet size, the 1 PAX ratio decreases slightly while the other ones increase. In fact, when more vehicles are available, the demand is greater; thus, the probability of finding further trip requests in an acceptable time or distance buffer from the actual ride(s) becomes higher. Therefore, the rides are more shared in the big fleet sizes and more seats are occupied. As illustrated in Fig. 7, 3 PAX ratio varies from 5 to 8% in the case of a standard 4-seats car and 6-seats minivan. However, the ratio of 4 PAX is less than 1%. Furthermore, the sixth seat of the minivan is almost never used. Actually, by comparing the other metrics one can observe that the differences between standard 4-seats car and 6-seats minivan are very small. In fact, given the amount of initial investment and operational costs of the bigger vehicles, the extra capacity may not necessarily be profitable. Therefore, it seems that standard 4-seats car is more compatible to the proposed service rather than a 6-seats minivan. Nevertheless, we have to keep in mind that the extra capacity could have an important impact on SAV's LoS. Furthermore, the 6-seats minivan can also be used for special requirements such as larger groups and families.

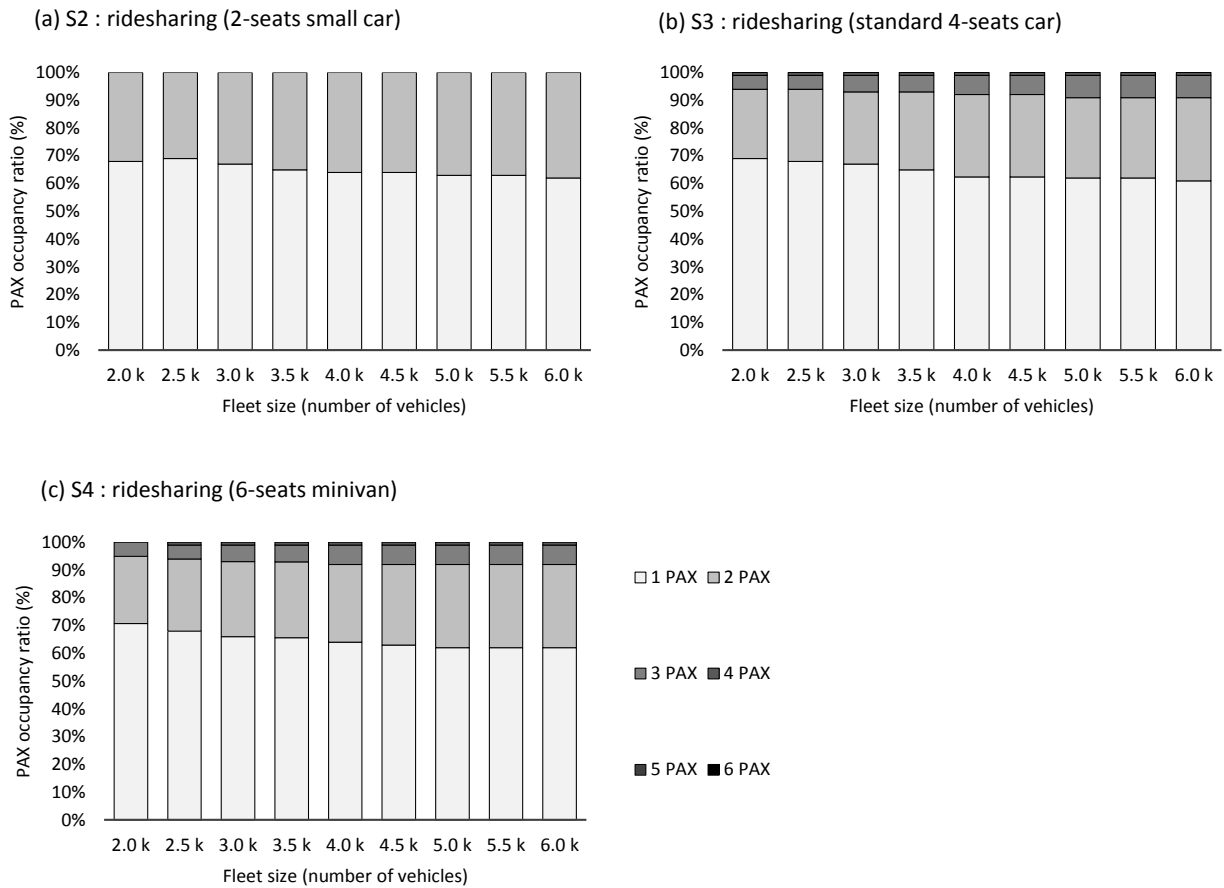

**Fig. 7.** On-board occupancy rates by the number of passenger for ridesharing scenarios.

The analysis on the origin and destination activities of trips performed by a fleet of 3.0 k standard 4-seats SAVs (Fig. 8) illustrates that almost half of all trips start from or end at users' homes. For the total of 9% of trips, the purpose at origin or destination is shopping or accompanying (escorting), indicating the importance of extra vehicle capacities in terms of the number of seats or luggage space. Also, given an important share of work and study activities at origin and destination (about 40%), it is likely that providing extra space and additional services for business, entertainment, and education purposes may provide a better customer experience while using SAV service.

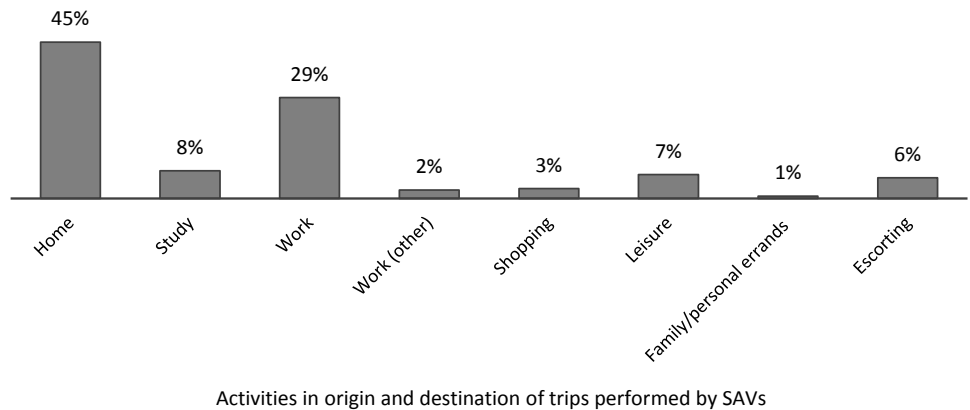

Activities in origin and destination of trips performed by SAVs

**Fig. 8.** The share of activities in origin and destination of trips performed by a fleet of 3.0 k standard 4-seats SAVs (percentage).

*4.5. Vehicle range*

Future SAVs are likely to be electric. In electric vehicles, the range is limited according to the battery capacity and specifications (e.g., weight and life cycle). In fact, given the important cost of battery among vehicles parts, its capacity may strongly affect the capital expenditure and operational expenses. According to the analysis of the SAV range for different scenarios and fleet sizes (Table 5), it is shown that the average driven distance of SAVs may intensely vary from 361 to 647 km. Furthermore, the vehicles do not have the same driven distance and for some vehicles, the average driven distance could be very long (even 975 km). This occurs when the demand is saturated and the vehicles are occupied for a long time during the day. By comparing the outputs, one can observe that the average driven distance correlates with the fleet usage ratio (Fig. 4) and the average number of rides per SAV (Table 4). In all fleet sizes, except when 2.0 k vehicles are simulated, the average driven distance of non-ridesharing SAVs is the largest compared to other scenarios. This result can be explained by the fact that SAV driven distances are shorter for the case of sharing the ride than when the ride is dedicated just to one passenger. However, in the case of the smallest fleet size, the fleet usage is dramatically lower compared to other fleet sizes due to the high expected waiting time and lower service request (Fig. 3). Thus, the average driven distance of non-ridesharing SAVs remains the lowest among all scenarios. The aforementioned distances could further increase by considering vehicle rebalancing, suggesting that vehicle ranges and possibly charging infrastructure need to be taken into account in future research.

15**Table 5**
Summary of vehicle driven distances for all scenarios and fleet sizes.

| Scenario | Fleet size | 2000 | 2500 | 3000 | 3500 | 4000 | 4500 | 5000 | 5500 | 6000 |
|---|---|---|---|---|---|---|---|---|---|---|
| S1- non-ridesharing | | | | | | | | | | |
| | Average driven distance (km) | 414 | 610 | 645 | 647 | 631 | 606 | 566 | 543 | 495 |
| | Maximum driven distance (km) | 652 | 975 | 907 | 894 | 895 | 866 | 925 | 822 | 884 |
| S2- ridesharing (2-seats small car) | | | | | | | | | | |
| | Average driven distance (km) | 469 | 549 | 528 | 543 | 503 | 477 | 422 | 408 | 377 |
| | Maximum driven distance (km) | 779 | 880 | 923 | 869 | 820 | 817 | 791 | 778 | 807 |
| S3- ridesharing (standard 4-seats car) | | | | | | | | | | |
| | Average driven distance (km) | 546 | 553 | 546 | 524 | 463 | 445 | 392 | 386 | 354 |
| | Maximum driven distance (km) | 820 | 919 | 866 | 900 | 825 | 866 | 799 | 797 | 691 |
| S4- ridesharing (6-seats minivan) | | | | | | | | | | |
| | Average driven distance (km) | 541 | 528 | 552 | 526 | 479 | 444 | 408 | 383 | 361 |
| | Maximum driven distance (km) | 861 | 847 | 888 | 870 | 779 | 883 | 761 | 841 | 714 |

*4.6. Ridesharing service cost*

The above-mentioned results are given when the cost of the ridesharing service is assumed to be 20% less than individual rides (0.4 Euro per kilometer compared to 0.5 Euro per kilometer for the individual ride). This price is encouraging enough for the travelers to prefer the ridesharing service to individual rides within the same fleet size according to the PKTs (Fig. 6). Although reducing service price and sharing rides lead to a higher PKT, the former may not be interesting for the operators as the service benefits for each kilometer of ride decrease, assuming that by an increase in PKT, the fixed cost of operation per kilometer remains unchanged. Thus, it is important to compare the benefit that an operator could gain due to the growth of PKT with the loss that occurs due to the reduction of profit per kilometer. In order to explore the evolution of service performance indicators, two lower prices for ridesharing services are assumed (i.e., 30% and 40% less than individual ride price) and the impacts on PKT and empty vehicle traveling distance in kilometer (EVK) are compared. As shown in Fig. 9, reducing service price results in 4-10% higher PKT compared to the initial ridesharing scenario, with the maximum value for 3.0 k SAVs in the second and third scenarios and 2.5 k SAVs in the fourth scenario. However, the EVK changes vary between -4% and 18% with the maximum values at the same fleet sizes in each scenario. As can be noticed, the reduction of service price does not cause proportionate improvements in the operational performance indicators of the major scenarios and fleet sizes. In fact, the increase of PKT as the main indicator of service profits, which is lower than 10% in the best case, is not enough to cover the loss of direct income occurred due to the lower service price (30-40%). In fact, the latter is certainly higher than 10% since the operational costs are included in the service price. Moreover, in the cases when an important growth of PKT occurs, the EVK increases and therefore the cost of service for operator grows, as well. Hence, for the fleet of 2.5 k to 3.5 k SAVs, the scenarios with initial service price remain more advantageous.

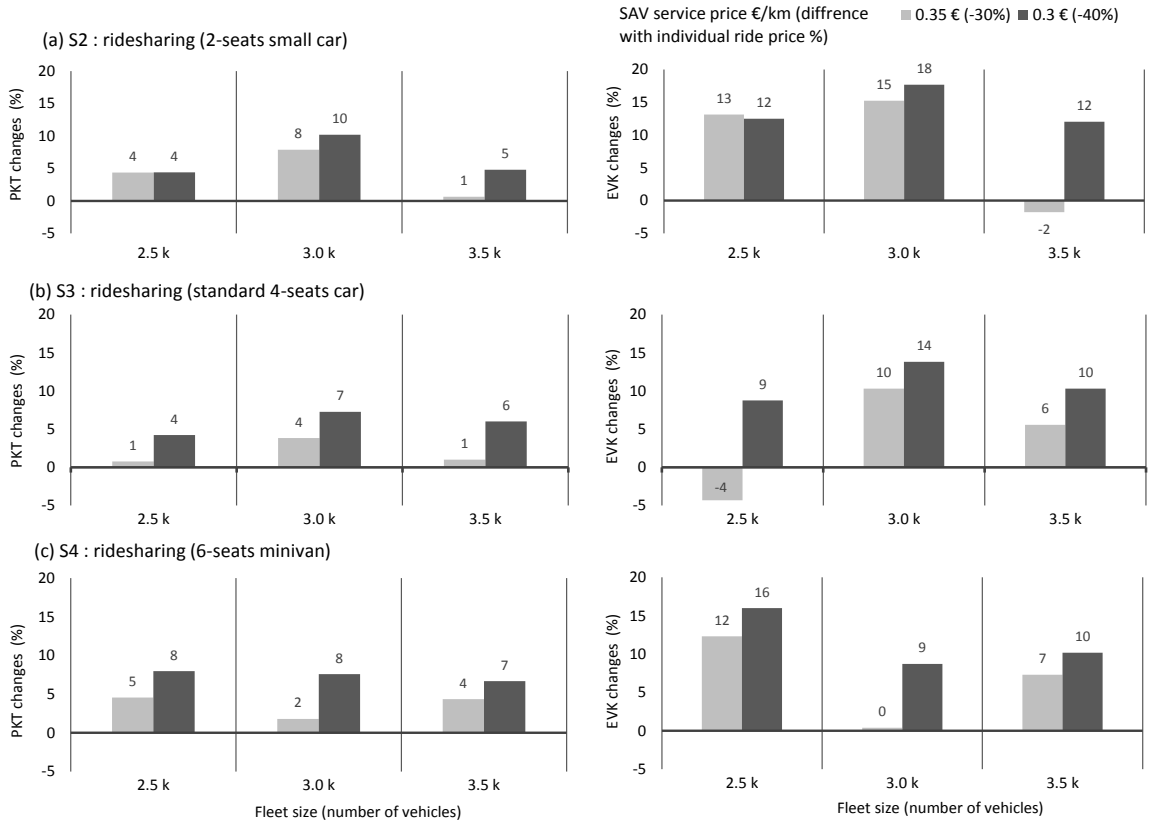

**Fig. 9.** The changes on PKT and EVK in the case of lower prices for ridesharing SAV services of 2.5 k to 3.5 k vehicles.

*4.7. Rebalancing strategy*

The fleet usage ratio and PKT may be improved by rebalancing SAVs. However, enabling this strategy can result in higher EVK. In order to explore the impacts on SAV service performances, the optimum fleet size of each scenario is re-simulated with rebalancing enabled. During these simulations, vehicles are reallocated to different cells with an area of 1 km$^2$ (used for demand aggregation) according to the cost flow minimization of idle vehicles and scattered requests. Empty vehicles are considered idle when there is no request after 10 minutes of stay. The reallocation process is done every 5 minutes. The costs of the single ride and ridesharing services are assumed to be as initial values (0.5 €/km for a single ride and 0.4 €/km for ridesharing). Table 6 illustrates the changes in performance metrics. As can be seen, modal share, fleet usage ratio, and in-vehicle PKT increase for all scenarios when SAVs are rebalanced. However, the empty distance ratio increases significantly. In fact, the growth of service benefits that is correlated with fleet usage and in-vehicle PKT occurs at the expense of extra operational costs due to empty traveling distance. Consequently, the decision on using a rebalancing strategy has to be made according to the cost and benefits that the operator of such services expects for each kilometer traveled by empty vehicles and passengers. Some other important changes occur in terms of the SAV's LoS. As presented in Table 6, the average waiting time has meaningfully decreased for the ridesharing scenarios after introducing rebalancing. This decrease occurs when the empty vehicles, which are far from the demand hubs, are reallocated to those zones. As a result, there are more vehicles available within lower wait times. Nevertheless, in-vehicle and detour times remain almost unchanged. Regarding in-vehicle time, the changes before and after introducing rebalancing strategy are minor since the trip patterns do not change significantly. Consequently, this indicator varies in the same order observed for various fleet sizes of the same scenario (Fig. 2). However, given the greater number of available vehicles and accordingly lower wait times, it seems that the detour time should similarly decrease. This decrease did not occur due to the higher demand as well as lower 1 PAX and bigger 2 and 3 PAX ratios. In fact, after enabling rebalancing, more rides are shared. Therefore, the average detour time remains almost unchanged. In the case of the individual ride scenario, since the service is saturated during morning and evening peak hours, the rebalancing strategy does not necessarily result in significant waiting time changes.

Similar to fleet usage and in-vehicle PKT, average and maximum driven distances increase for all scenarios. In other words, by introducing a rebalancing strategy the vehicles need to have larger batteries or need to recharge more frequently.

Table 6
Performance metrics' changes before and after enabling the rebalancing strategy.

| Scenario | Non-ridesharing (3.5 k) | | Ridesharing 2-seats small car (2.5 k) | | Ridesharing standard 4-seats car (3.0 k) | | Ridesharing 6-seats minivan (3.0 k) | |
|---|---|---|---|---|---|---|---|---|
| | no rebalancing | with rebalancing | no rebalancing | with rebalancing | no rebalancing | with rebalancing | no rebalancing | with rebalancing |
| SAV modal share (%) | 6.0 | 6.3 | 4.6 | 5.2 | 5.3 | 6.4 | 5.4 | 6.4 |
| Average waiting time (min) | 18.5 | 18.4 | 18.9 | 13.9 | 20.7 | 13.1 | 21.1 | 14.8 |
| Average in-vehicle time (min) | 38.5 | 38.7 | 43.9 | 44.1 | 46.0 | 45.2 | 46.0 | 44.8 |
| Average detour time (min) | N/A | N/A | 4.7 | 5.2 | 6.1 | 5.8 | 6.0 | 5.9 |
| Fleet usage ratio (%) | 59 | 68 | 50 | 66 | 50 | 66 | 51 | 67 |
| Empty distance ratio (%) | 14 | 20 | 14 | 26 | 15 | 24 | 16 | 24 |
| In-vehicle PKT (km) | 1.93 M | 2.08 M | 1.53 M | 1.81 M | 1.97 M | 2.40 M | 1.97 M | 2.41 M |
| 1 PAX ratio (%) | 100 | 100 | 69 | 63 | 67 | 59 | 66 | 61 |
| 2 PAX ratio (%) | N/A | N/A | 31 | 37 | 26 | 33 | 27 | 31 |
| 3 PAX ratio (%) | N/A | N/A | N/A | N/A | 6 | 7 | 6 | 7 |
| 4 PAX ratio (%) | N/A | N/A | N/A | N/A | 1 | 1 | 1 | 1 |
| 5 PAX ratio (%) | N/A | N/A | N/A | N/A | N/A | N/A | <1 | <1 |
| 6 PAX ratio (%) | N/A | N/A | N/A | N/A | N/A | N/A | 0 | 0 |
| Average driven distance (km) | 647 | 746 | 549 | 715 | 546 | 707 | 552 | 723 |
| Max. driven distance (km) | 894 | 978 | 880 | 964 | 866 | 896 | 888 | 939 |

## 5. Discussion and conclusion

The rising popularity of carsharing and technological advancements on electric and autonomous vehicles has led to the emergence of new shared mobility systems. Some car manufacturers and transportation network companies (e.g., Uber and Lyft) have already announced their plans for deploying SAVs in the future. Understanding dynamic tradeoffs between service configuration and demand is an important prerequisite for delivering such services. This study sought to investigate the design of an SAV service considering its demands responsive to the network, user taste variations, and traffic in a multi-modal context. Simulations of various SAV fleet sizes and capacities considering ridesharing and rebalancing strategies across the Rouen Normandie metropolitan area in France provide initial insights. As suggested by these simulations, the SAVs performance is strongly correlated to the fleet size, specifically in the case of individual ride service. The results show that the SAV modal shares vary from 3.1% to 7.6% for different fleet sizes of 2.0 k to 6.0 k vehicles. While the SAV modal share of small fleet size for the individual ride is the minimum among all scenarios including ridesharing with various vehicle capacities, this term is the greatest for the medium and big fleet sizes. The latter actually occurs when the fleet of individual ride service exceeds a critical size (i.e., 2.5 k), from which the smaller fleet size results in a significant decline of fleet usage (i.e., 34% compared to 8% that is observed from 3.0 k to 2.5 k vehicles). In fact, for fleet sizes this small, the expected wait times increase meaningfully and lead to very low service utility compared to other alternatives; and the demand decreases, accordingly. On the contrary, once the peak hour demand is satisfied in acceptable wait times, the service performance decreases slightly by increasing fleet size. This variation is not as significant as in ridesharing scenarios where the change of fleet size has less important impacts on the LoS. The results also suggest that the average waiting time, which is estimated when the interaction of demand and supply is relaxed (and therefore it is different from the aforementioned expected waiting time), decreases meaningfully for the smallest fleet sizes. This decrease is actually contrary to what is usually suggested when the SAV simulation with predefined or static demand is performed and shows the importance of considering dynamic demand in the simulation of on-demand services.

Further analysis reveals that in the case of ridesharing services without vehicle rebalancing, the changes in average waiting time remain insignificant for the medium and big fleet sizes. Nevertheless, this indicator decreases in the individual ride service for the big fleet sizes. In fact, in ridesharing scenarios since the service is never saturated in the case of medium and big fleet sizes, when an upcoming request is registered, the nearest vehicle with available seats is allocated to it. However, in individual ride services, the nearest empty vehicle has to be allocated to that request. Bigger fleet sizes result in a large number of vehicles available in individual ride services and the waiting time decreases accordingly. The increase in fleet size similarly results in more available seats in ridesharing scenarios (up to 5%). However, this change is not significant since the fleets are not as occupied as in individual ride scenario and the increase in seat availability does not have an important impact on the availability of service. It is also shown that average in-vehicle and detour times vary slightly according to the fleet size. These changes are relatively minor (less than 4 minutes) and follow almost the same trend.

18By comparing the fleet usages and empty distance ratios of different scenarios, we found that the optimum fleet sizes for the individual ride and ridesharing cases are different. These results suggest that while the best fleet size of individual rides is 3.5 k, in the case of a small car with two seats, 2.5 k vehicles and for standard 4-seats car and 6-seats minivan, 3.0 k vehicles are the best performing fleet sizes. Based on the obtained results, considering transport related service performance, there are no big differences between standard 4-seats car and 6-seats minivan. Since the bigger capacity vehicle may be financially less efficient due to the higher vehicle and operational costs, it seems that 6-seats minivan cannot be a performant alternative. Nevertheless, we have to bear in mind that the extra capacity and seats may potentially affect the user comfort perception and consequently their choice in the real world. Further comparison of four suggested fleet sizes illustrates that given the relatively high in-vehicle PKT and low empty distance ratio of the 3.0 k vehicles with share rides, this scenario is the best option among all the considered scenarios. Furthermore, taking into account the trade-off between waiting and detour times and service cost, we consider that the proposed pricing scheme for SAV ridesharing service (20% less compared to individual ride) is attractive enough for users. In addition, the results show that a decrease in ridesharing service prices up to 40% of the individual ride does not cause proportionate improvements on the operational performance indicators and is not beneficial for the operator.

Importantly, enabling vehicle rebalancing is found to have a profound effect on both user and service-related metrics. For optimum fleet sizes of ridesharing scenarios, rebalancing leads to shorter average waiting times (i.e., 25-35%). However, in individual ride scenario, this indicator remains unchanged since the service is already saturated in peak hours without enabling rebalancing. Although in-vehicle PKT and empty distance ratio increase for all scenarios, the change in the latter indicator is relatively more important (i.e., 42-85% against 7-17%). Besides, the detour times remain almost unchanged. Given these indicators, the decision on enabling rebalancing strategy should be made according to the financial analysis based on the cost and benefits that operator of such services expects for each kilometer traveled by passengers and empty vehicles.

The average driven distances for optimum scenarios without rebalancing varies from 546 to 647 km. Given the relatively lower range of today's electric vehicles, it will be necessary to recharge the majority of SAVs during the day. Furthermore, enabling rebalancing leads to longer average traveled distance (i.e., 707 to 746 km). Considering the maximum driven distance that exceeds even 900 km, these indicators suggest that future SAVs will necessarily require some recharging infrastructure.

The employed agent-based simulations in this study incorporate users' trip pattern and taste variation developed previously by the authors. The detailed service usage indicator estimated for 1-h time slices illustrates a shift of the morning peak-hour service usage from 7-9 a.m. for the small fleet sizes to 8-10 a.m. for big fleet sizes. This result indicates the effect that different SAV users with different trip patterns and taste variation may have on the service usage and prove the importance of considering that user differentiations in SAV demand modeling and simulation.

While these investigations and results offer a broad and new understanding of SAV service performance and design, there are several opportunities for improvement. For example, rather than having the same pricing scheme for all rides, future efforts should examine dynamic pricing (e.g., time-based or demand-based). In addition, an improved rebalancing strategy may be proposed and evaluated within this simulation framework. The authors further plan to integrate electric SAV (ESAV) into the simulation and investigate the evolution of service performance indicators according to various charging station positioning and vehicle ranges.

## Acknowledgments

This research work has been carried out in the framework of IRT SystemX, Paris-Saclay, France, and therefore granted with public funds within the scope of the French Program "Investissements d'Avenir". The authors would like to thank Groupe Renault for partially financing this work and Métropole Rouen-Normandie for providing the data.

This research work has been carried out in the framework of IRT SystemX, Paris-Saclay, France, and therefore granted with public funds within the scope of the French Program "Investissements d'Avenir". The authors would like to thank Groupe Renault for partially financing this work and Métropole Rouen-Normandie for providing the data.